\documentstyle[11pt]{article}
\begin{document}
\title{
ON THE SPACE REFLECTIONS DEFINITION PROBLEM IN THE
MAGNETIC CHARGE THEORY}
\author {E.A.TOLKACHEV, L.M.TOMILCHIK\\
  {\it Institute of Physics, Academy of Sciences of Belarus}\\
{\it Minsk 220072, Belarus}\\
\and YA.M.SHNIR{\thanks {Supported by the Alexander von Humbold
Foundation; Permanent address: 
Institute of Physics, Academy of
Sciences of Belarus, Minsk, Belarus}}\\
{\it Department of Mathematics,}\\
{\it Technical University of Berlin, Germany}
}
\maketitle
\begin{abstract}
A new way to define the operation of P-inversion in the theory with
a magnetic charge is presented.
\end{abstract}

At this conference, dedicated to memory of Professor A.O.Barut,
it is seemed to be pertinent to consider in a rather retrspective view
some fragments of his rich scientific legacy. Ecseptionally active
scientific and human approachability by Professor Barut was his modus
vivendi. He always was open for collaboration.

The $P$-invariant quantum mechanical dichotomic model of charge-monopole
interaction \cite{Barut} proposed by Professor Barut more than twenty years
ago is demonstrated below as an example his original ideas influence upon
our investigations. Later this influence was developed in a very fruitful
direct scientific cooperation and in the same time in imforgetable
personal contacts.

The problem  of P invariance of the electrodynamics with electric and
magnetic charges has a very long story (see, e.g.\cite{Sach} - \cite{3}).
Despite the quite definite answer has been obtained in \cite{4,5}
there still are some new endeavours to get a progress in this
direction. The gist of the matter can be easily be demonstrated on the
example of classical field theory.
\par
Let us consider the pure SU(2) gauge model with the Lagrangian
$$ L = -{1\over 4} F^a_{\mu, \nu} F^{a \mu, \nu},\eqno(1)
$$
where the field tensor is
$$ F^a_{\mu, \nu} = \partial_{\mu }A _{\nu}^a - \partial_{\nu }
A _{\mu}^a
- e {\varepsilon}_{abc}A _{\mu}^b A _{\nu}^c, ~~~~~~(a = 1,2,3),
\eqno(2)
$$
and $A _{\mu}^a $ is the gauge potential.

As is well known the corresponding equations of motion have the
Wu-Yang solution
$$
A _k^a = {\varepsilon}_{a k m}~{r_m \over r^2}.
\eqno(3)
$$
It is obvious that the model is invariant under the reflection
${\bf r} \to -{\bf r}$ and the Wu-Yang potential is transformed as a vector:
$A _k^a ({\bf r}) \to -A _k^a ({\bf r})$. But on the other hand there
is an abelization of the solution (3) by means of the gauge
transformation
$$
S A _k^a ({\bf r}) S^{-1} + {i\over e} ({\nabla}_k S) S^{-1} =
A_k^D ({\bf r,n}) \sigma _3,
\eqno(4)
$$
where
$$ S = \left(
\begin{array}{cc}
\cos {\theta \over 2} & -\sin {\theta \over 2}
e^{-i\phi} \\[3pt] 
\sin {\theta \over 2}
e^{i\phi} & \cos {\theta \over 2} 
\end{array} \right),
\eqno(5)
$$
and
$$
A_a^D ({\bf r,n}) =
{\varepsilon _{abc} r_b n_c \over {r - (r_a n_a)}}
$$
is the Dirac potential.

Of course,  the gauge transformation (4) can not violate the P invariance of
the theory. This point gives a possibility to obtain a new way  to define the
operation of P inversion in the  theory with a magnetic charge.

Indeed, under the P-inversion
$\theta \to \pi - \theta, \quad \phi  \to \phi +  \pi$ equation (4) is
transformed as
$$
S_p  (-A _k^a ) S_p^{-1} - {i\over e} ({\nabla}_k S_p) S_p^{-1} =
A_k^D ({\bf -r,n}) \sigma _3,
$$ 
or
$$
 S_p  A _k^a  S_p^{-1} + {i\over e} ({\nabla}_k S_p) S_p^{-1} =
- A_k^D ({\bf -r,n}) \sigma _3,
\eqno(6)
$$
where
$$ S_p = 
\left(
\begin{array}{cc}
\sin {\theta \over 2} & \cos {\theta \over 2}
e^{-i\phi} \\[3pt]
 -\cos {\theta \over 2}
e^{i\phi} & \sin {\theta \over 2} \end{array} \right).
\eqno(7)
$$
We have to transform Eq.(6) to the (4) if we like to check the P invariance of the
model.  To this end we write
$$S_p = R S, \quad {\rm i.e.}\quad R = S_p S^{-1} = \pmatrix{0 & e^{-i\phi} \cr
-e^{i\phi} & 0} =\pmatrix{0 & 1 \cr 1 & 0}
\pmatrix{-e^{i\phi} & 0 \cr 0 & e^{-i\phi} }
$$

Then, bring  the matrix of transformation R from the left to right in Eq.(6) we
really get the Eq.(4) because the matrix $\pmatrix{0 & 1 \cr 1 & 0} $ changes the
common sigh of the r.h.s. and the matrix $\pmatrix{-e^{-i\phi} & 0 \cr 0 &
e^{i\phi} }$ changes the sigh of ${\bf r}$. Thus the r.h.s. of (4) is a true
vector up to a gauge transformation determined by $R^{-1}$. Consequently the
Dirac potential ${\bf A}^D({\bf r},{\bf n})$ is a pseudovector up to $U(1)$
gauge transformation determined by $\exp(i\phi)$. The central point of our
cosideration is transformational properties of matrix $S$ under spatial
reflection. Let ${\bf A}_k^a({\bf r})$ in eq. (4) be any true vector, then
 the corresponding r.h.s. is a vector up to the gauge transformation
 determined by $S\,S_{\gamma}^{-1}$. It is worth noting that in quantum
theory the use of the unitary transformations determined by (5) leads also
to redefinition of the P-inversion operator
$$ P' = S P S^{\dagger} = S (S_p)^{-1} P = R P.
 \eqno(8)
$$

Obviously, $P'$ is a Hermitian operator, i.e.
$$ (P')^{\dagger} = P (R^{-1})^{\dagger} = R^{-1} P.
$$

The matrix R up to nonsignificant here factor $\sigma _3 $ concided
with the matrix
$$ {\hat R} = \pmatrix{0 & e^{-2i\mu\phi} \cr
-e^{2i\mu\phi} & 0}, \qquad \mu = eg = 1/2,
 \eqno(9)
$$
\noindent introduced in \cite{6}.
It is easy to check that using the combination of the standard
reflection of the space coordinates and the gauge transformation
(9) allows to restore formal P invariance of the model with Barut's
dichotomic  Hamiltonian \cite{Barut}, \cite{8}
 $$ H = H({\bf A}^D \sigma_3) $$
But we seems to run into a slight problem if we attempt to include other external electromagnetic 
fields ${\bf A}^e$ into this framework together with ${\bf A}^D$. Indeed, if we are to follow the usual embedding
procedure all the electromagnetic potentials have to be multiplied by $\sigma _3$. Only in this           
case is this model effectively equivalent to the U(1) model with the standard rule of adding fields.
In this case the non-relativistic Hamiltonian operator of the charge-dyon system would be
$$ H = - \frac {1}{2M} ({\bf P} + (e {\bf A}^D + e{\bf A}^e) \sigma_3 )^2  - e A_0^D  \sigma_3 - e A_0^e  \sigma_3 .
\eqno(10)
$$
Instead we can write a P-invariant Hamiltonian operator as 
$$ H = - \frac {1}{2M} ({\bf P} + e {\bf A}^D \sigma_3 + e{\bf A}^e I )^2  - e A_0^D  \sigma_3 - e A_0^e   I.
\eqno(11)
$$

This, as a matter of fact, implies an extension of electrodynamics because the model with Hamiltonian
(11) is invariant under the gauge transformation with $U(1) \otimes U(1)$ group  and describes the interaction 
of the quantum particle carrying pseudoscalar $(e \sigma_3)$ and scalar $(e I)$ charges with pseudovestor 
$( A_0^D,  {\bf A}^D)$ and vector $( A_0^e,  {\bf A}^e)$ fields, correspondingly. Moreover, with this 
definition it is not necessary to fix the same interaction constant $e$. Indeed, because the eigenfunction
 $\Psi$
of the Hamiltonian operator (11) is a two-component entity, the natural gauge transformation is 
$\Psi\to\exp\{i(e+e' \sigma_3)\} \Psi$.

Let us illustrate these arguments in the example ot the calculation of selection rules for dipole radiation.

It is easy to see that the dipole moment operator corresponding
to the Hamiltonian (10) is $e({\bf r} {\sigma}_{3})$
and the corresponding operator  for
the Hamiltonian (11) s $e({\bf r} I)$. In the first  case   the  total  set
includes the charge operator $e\sigma _{3}$, therefore a matrix element  should
be calculated between the following type of wavefunctions:
$$
\pmatrix{\Phi_{Njm\mu }({\bf r})\cr 0},\hspace{8pt} {\rm or}
\hspace{8pt} \pmatrix{0\cr \Phi_{Njm\mu }({\bf r})},$$
\noindent where $\mu = eg$,  $\Phi_{Njm\mu}({\bf r}) = R_{Nj}(r)Y_{jmq}(\theta,
\varphi)$  are  the  wavefunctions  of  the
electrically charged particles in the field ${\bf A}^D $,  $R_{Nj}(r)$  is  their
radial part, $j$ and $m$ are the quantum numbers corresponding to the eigenvalues of the operator 
${\bf J}^2$ and $J_3$ (${\bf J} $ is the operator for the total angular momentum of the system) and the 
generalized spherical harmonics $ Y_{jmq}(\theta,
\varphi)$ are expressed through the standard Jacobu polynomial 
$P_n^{(\alpha,\beta)}(x)$\cite{Tamm}:
$$Y_{jmq}(\theta, \varphi)  = N (1 -x)^{-(m + \mu)/2} (1 + x)^{-(m - \mu)/2} P_{j+m}^{(-m-\mu), (-m + \mu)}(x)
e^{ i(m + \mu)}\varphi,
\eqno(12)
$$
\noindent where $x = \cos \theta$, and 
$$N  = 2^{m}\sqrt{{(2j+1)(j-m)!(j+m)!}\over {4\pi (j-\mu)!(j+\mu)!}}.$$
After integration
$$
\int d^{3}x {\Phi}_{Njm\mu}^* ({\bf r}) {\bf r}
{\sigma}_3  {\Phi}_{Njm\mu}({\bf r}),
\eqno(13)$$  
we obtain the selection rules that coincide with results of \cite{4}:
$$
\Delta j = 0,\pm 1,\hspace{12pt} \Delta m = 0, \pm 1\eqno(14)
$$
\noindent Thus, the parity violating transitions  with $\Delta j = 0$
are  allowed together with the standard transitions with $\Delta j = \pm 1.$

In the  second  case  however,  among  operators  commuting  with
Hamiltonian  (11) there  are  the  operator $\sigma _3$  and  generalized $P$
inversion operator (8), which however  do  not  commute  with  each
other. Choosing the general eigenfunctions of the operators  (8) and (11) as 
$$
{\Psi}_{Njm\mu}({\bf r}) = \pmatrix{{\Phi}_{Njm\mu}({\bf r})\cr
{\Phi}_{Njm-\mu}({\bf r})},\eqno(15)
$$
\noindent we calculate the matrix element of the dipole moment operator
$$
\int d^{3}x {\Psi}_{Njm\mu}^{*}({\bf r}){\bf r} I
{\Psi}_{Njm\mu}({\bf r}),\eqno(16)
$$
\noindent and integrate over the angular part of (16). We have for instance:
$$
\int^{}_{\Omega}d\Omega {\Psi}^{*}_{j^{\prime}m^{\prime}{\mu}^{\prime}}
(\theta ,\varphi) \cos \theta  I {\Psi}_{j m \mu }(\theta ,\varphi ) =
$$
$$ C\Biggl(\int^{}_{\Omega }d\Omega {Y}_{j^{\prime}m^{\prime}{\mu}^{\prime}}
(\theta ,\varphi)Y_{100}(\theta ,\varphi )Y_{j m \mu }(\theta ,\varphi )
$$
$$
\pm  \int^{}_{\Omega }d\Omega Y_{j^{\prime}m^{\prime}{\mu}^{\prime}}
(\theta ,\varphi )Y_{100}(\theta ,\varphi ){Y}_{j m -\mu}(\theta,\varphi)\Biggr)
$$
$$={C^{\prime}}{\pmatrix{j^{\prime}& 1 &j \cr -m & 0 & m}}
\Biggl[ {\pmatrix{j^{\prime} & 1 & j \cr
-\mu & 0 & \mu}}-{\pmatrix{j^{\prime}&1&j\cr \mu & 0 & -\mu }}\Biggr],\eqno(17)$$
\noindent where $C$ and $C^{\prime}$ are some constants.
Taking into account the  familiar
properties of the $3-j$ symbols
$$
{\pmatrix{j^{\prime}&1&j\cr -\mu & 0 & \mu}}  = (-1)^{j^{\prime} + j + 1}
{\pmatrix{j^{\prime}&1&j\cr \mu & 0 & -\mu}}$$
\noindent we see that the integral (17) is non-zero only for
$\Delta j = \pm  1.$  In  an
analogous  way,  after  integration  of  the  angular  integrals  of
$\sin \theta e^{\pm i\varphi }$ we obtain the selection rules:
$$\Delta j = \pm 1,\hspace{12pt} \Delta m = 0, \pm 1. \eqno(18)$$
\noindent that is to say the dipole  transitions  with  parity  violation  are
absent in this case.

Let us note that the use of the wavefunction (15)  in  (13)  does
not modify the selection rules (14) but the value  of  the  integral
(13) in this case is increased twofold.

Thus  the  restoring  of  the  standard  selection  rules  and $P$
invariant description of this system are achieved by  means  of  the
gauge group extension to $U(1)\otimes U(1)$.At the same time  the  hypothesis
about the Abelian magnetic charge is not connected with an extension
of the symmetry group but based on the transition to the non-trivial
fiber-bundle over the space-time base with the structure group $U(1)$
when the connection (potential) and sections (wavefunctions)  cannot
be described globally.\\[2mm]

{\bf Acknowledgements}\\[2mm]

L.M.Tomilchik is deeply indebted to Prof. I.H.Duru
for the kind invitation to Barut Memorial Conference on Group Theory in Physics
and would like to thank him for hospitality at the ICPAM.

\end{document}